# From Quantum Information to Gravitation (in German)

Thomas Görnitz

FB Physik, Goethe-Universität

Frankfurt/Main

goernitz@em.uni-frankfurt.de

> …that it may be no more appropriate
> to quantize the Einstein equation
> than it would be to quantize
> the wave equation for sound in air.
> Ted Jacobson

## Abstract

To unite quantum theory and general relativity, in a new access it is shown that from a theory of an abstract quantum information – called Protyposis – the theory of general relativity can be deduced by means of few and physically good founded reasons. "Abstract" quantum information means that primarily no special meaning is connected with it. The resulting cosmology has an isotropic and homogeneous metric and solves the so-called cosmological problems. For the Protyposis it follows an equation of states for energy density and pressure that fulfils all the energy conditions and that also gives an explanation for the "dark energy". If it is demanded that the relations between spacetime structure and the material contend also hold for deviations from this ideal cosmology than General relativity results as a description for local inhomogenities.

-----

Um Quantentheorie und Gravitation zu verbinden, wird in einem neuen Zugang gezeigt, dass aus einer Theorie einer abstrakten, d.h. einer primär noch bedeutungsfreien Quanteninformation – als Protyposis bezeichnet – mit wenigen und physikalisch gut begründbaren Annahmen eine realistische Kosmologie folgt. Diese besitzt eine homogene und isotrope Hintergrundmetrik und löst die sogenannten kosmologischen Probleme. Für die Protyposis folgt weiterhin eine Zustandsgleichung für Energiedichte und Druck, die alle Energiebedingungen erfüllt und die auch eine Erklärung für die Erscheinungen der „dunklen Energie" liefert. Die Allgemeine Relativitätstheorie ergibt sich dann als eine Beschreibung lokaler Abweichungen von dieser idealen Kosmologie, wenn für die Abweichungen vom homogenen und isotropen Zustand weiterhin die Gültigkeit der Beziehungen zwischen Materie und Raumzeitstruktur gefordert wird, die für den Kosmos als Ganzen gilt.

# Einleitung

Seit vielen Jahren wird intensiv daran gearbeitet, eine Quantisierung der Gravitation zu erreichen. Bisher ist diesen Versuchen allerdings kein durchschlagender Erfolg beschieden worden. Ausgangspunkt der Überlegungen war dabei bis jetzt die Allgemeine Relativitätstheorie – eine Struktur höchster mathematischer Schönheit und Komplexität. Sie wurde als Grundlage gewählt, um sie dann – analog zu anderen Feldtheorien – auf die eine oder andere Weise zu quantisieren.

Das Tor von der Relativitätstheorie zur Quantentheorie hat sich bisher nicht öffnen lassen. Wenn man einen Schritt zurücktritt, so wird erkennbar, dass es möglich ist, das Tor in der anderen Richtung zu öffnen – von der Quantentheorie hin zur Gravitation.

Einige weitere Gründe sprechen für eine Änderung des bisherigen Verfahrens. Sie haben auch einen erkenntnistheoretischen Aspekt und beziehen sich auf Voraussetzungen, die bisher wenig bedacht wurden. Daher müssen sie über lediglich mathematische Überlegungen hinausreichen.

Aus einer Theorie einer abstrakten, d.h. einer primär noch bedeutungsfreien Quanteninformation folgt mit wenigen und physikalisch gut begründbaren Annahmen zum einen eine realistische Kosmologie mit einer homogenen und isotropen Hintergrundmetrik, zum anderen eine Zustandsgleichung für Energiedichte und Druck dieser Quanteninformation.

Diese Kosmologie steht heute in wesentlich besserer Übereinstimmung mit den Beobachtungsdaten als bei ihrer ersten Vorstellung im Jahre 1988.[1] Sie löst die sogenannten kosmologischen Probleme und liefert unter anderem eine Erklärung für die Erscheinungen der „dunklen Energie".

Aus dieser auf abstrakte Weise quantentheoretisch begründeten Kosmologie ergeben sich die Einsteinschen Gleichungen, wenn für eine Beschreibung von lokalen Abweichungen von dem homogenen und isotropen Zustand die weitere Gültigkeit der für den Kosmos gefundenen Beziehungen zwischen Materie und Raumzeitstruktur gefordert wird.

Dieser Weg bricht mit einigen der bisher als unumstößlich angesehenen erkenntnistheoretischen Postulaten aus dem Umfeld der Allgemeinen Relativitätstheorie. Jedoch bleibt die fundamentale Bedeutung der aus ihr entwickelten Näherungen als die gegenwärtig beste Beschreibung für alle lokalisierten Gravitationsphänomene davon unbeeinflusst.

# Beziehungen zwischen Allgemeiner Relativitätstheorie, Quantentheorie und Kosmologie

Einstein schreibt über die Allgemeine Relativitätstheorie: „Sie gleicht aber einem Gebäude,

---
[1] Görnitz 1988 a, b

dessen einer Flügel aus vorzüglichem Marmor (linke Seite der Gleichung), dessen anderer Flügel aus minderwertigem Holze gebaut ist (rechte Seite der Gleichung)."[2] Nun besteht aber der Sinn einer Gleichung darin, dass beide Seiten gleich sind. Wenn also die rechte lediglich aus Holz besteht, so gilt naturgemäß das gleiche für die linke Seite. Die Wahrnehmung des geschilderten Unterschiedes ist also offenbar nicht wirklich zutreffend.

Natürlich sind die geometrischen Beziehungen der Einsteinschen Gleichungen von einer unübertrefflichen Klarheit, aber die eigentliche physikalische Fragestellung – im Gegensatz zu einer rein mathematischen – besteht darin, die Beziehungen zur Empirie herstellen zu können.

Seit der Endeckung der Quantentheorie kann man wissen, dass die wunderbaren mathematischen Modelle von Raum-Zeit-Mannigfaltigkeiten und Geodäten bei einer sehr genauen Betrachtung, die dann naturgemäß die Quantentheorie einschließen muss, hinfällig werden. Daher erscheint es notwendig, den vielfach versuchten Übergang von der Gravitation zu den Quantenphänomenen zu hinterfragen, der bisher in dieser oder jener Form für eine Quantisierung der Allgemeinen Relativitätstheorie angenommen wurde.

Die Quantenstrukturen sind der Physik erst aufgezwungen worden, als sie eine sehr große Genauigkeit der Naturbeschreibung erreicht hatte. Erst bei sehr großer Genauigkeit werden sie bedeutsam. Daraus kann man schließen, dass keineswegs jede mathematische Struktur, welche die Naturvorgänge hinreichend gut modelliert, auch eine quantisierte Form zulassen muss. Lediglich diejenigen mathematischen Modelle, die als fundamental anzusehen sind und die besonders gut an die Natur angepasst sind, werden sich quantisieren lassen.

Von dieser Überlegung ausgehend soll nun die Allgemeine Relativitätstheorie betrachtet werden. Ganz allgemein kann man davon sprechen, dass der Bedeutungsgehalt einer Gleichung äquivalent ist zur Menge aller ihrer Lösungen. Mit der Gleichung hat man – zumindest im Prinzip – die Menge ihrer Lösungen, und aus sämtlichen Lösungen lässt sich – wiederum zumindest im Prinzip – die Gleichung rekonstruieren.

Im Gegensatz zu den sonstigen Gleichungen der Physik zeichnen sich die Einsteinschen Gleichungen nun dadurch aus, dass eine Lösung – falls sie nicht lediglich als eine Näherung verwendet wird – stets einen ganzen Kosmos in seiner vollen raumzeitlichen Erstreckung beschreibt.

Wenn man Physik betreiben möchte und nicht reine Mathematik, so ist immer eine Anbindung an die Empirie notwendig. Aus Sicht der Physik kann der Kosmos definiert werden als die

---

[2] Einstein, 1984, S. 90

Menge aller der Erscheinungen, für die eine Kenntnisnahme nicht prinzipiell unmöglich erscheint. Der Kosmos ist also die Gesamtheit all der möglichen physikalischen Empirie. Wenn man dies akzeptiert, so ist der Kosmos als Gegenstand der Physik notwendig ein einziger. Eine Mehrzahl davon ist einsichtigerweise im Rahmen der Physik unmöglich.[3]

Da die Einsteinschen Gleichungen unendlich viele Lösungen besitzen, von denen aber höchstens eine den realen Kosmos unserer physikalischen Empirie beschreiben kann, während alle anderen mit der Empirie nichts zutun haben können, muss mit diesen Gleichungen notwendig vieles behauptet werden, was über die Physik hinaus in die reine Mathematik gehört.

Diese Schlussfolgerung ist einer der Gründe, die gegen eine Quantisierung der Allgemeinen Relativitätstheorie sprechen.

Andererseits sind die experimentellen und observatorischen Überprüfungen der Allgemeinen Relativitätstheorie – bzw. von linearisierten Näherungen von ihr – so gut, dass kein Versuch gerechtfertigt werden kann, diese Gleichungen nicht als eine sehr gute Beschreibung der Beobachtungen und der Erfahrungen anzusehen, die aber trotzdem als Näherung verstanden werden muss.

Das Ziel einer Verbindung von Kosmologie und Physik betrifft aber noch weitere Aspekte, die nur selten verbalisiert werden.

**Das Empirieproblem**

Selbstverständlich spricht nichts dagegen, Kosmologie als einen Teil der reinen Mathematik zu verstehen und kosmologische Lösungen der Allgemeinen Relativitätstheorie in all ihren interessanten mathematischen Eigenschaften zu untersuchen, ohne damit einen Anspruch zu verbinden, dass diese Ergebnisse etwas mit der beobachtbaren Realität zu tun haben sollen. Andererseits verzichten viele Physiker und Astronomen auf kosmologische Modelle – die definitionsgemäß den Kosmos als Ganzen betreffen – und beschränken sich auf die Betrachtung der Metagalaxis. Dieser Ausschnitt wird aber nur dann mit Bedeutung für das Ganze sein, wenn er bereits einen wesentlichen Teil des gesamten Kosmos erfasst und das Unbeobachtbare nur einen geringen Rest darstellt. Eine solche kritische Haltung gegenüber der Kosmologie wurde beispielsweise von C. F. v. Weizsäcker vertreten. Er berichtete darüber, dass er auf einer

---

[3] Allerdings kann man natürlich in Mathematik, Science Fiction oder Esoterik einen Plural von Universen bilden. Damit tritt man aber aus dem Rahmen der Physik heraus, denn das einzige, was man von solchen hypothetischen anderen Universen wissen kann – dies allerdings mit absoluter Gewissheit – ist, dass jede empirische Kenntnis über sie unmöglich ist.

Konferenz Stephen Hawking gefragt habe, ob es denn den Kosmos „als Objekt" überhaupt geben würde. Weizsäcker meinte damit die sehr berechtigte Frage, wie sinnvoll es ist, das „Unikat Kosmos" als spezielle Lösung eines „allgemeinen Gesetzes" und als „Objekt für uns" zu behandeln. Hawking fasste wahrscheinlich die Frage so auf, als ob Weizsäcker an der Realität der Gegenstände der Astrophysik zweifeln würde. Solches war natürlich überhaupt nicht Weizsäckers Meinung, aber seitdem war, so Weizsäcker, der Gesprächsfaden zwischen ihnen beiden abgeschnitten.

Wenn man also die Kosmologie als einen Teil der Physik betrachten möchte, dann muss man sich der Frage der empirischen Relevanz der Beobachtungsdaten stellen.

Hiermit entsteht das Empirieproblem.

Neben den anderen kosmologischen Problemen, wie z.B. den sogenannten Flachheits- und Horizontproblem oder dem Problem der sogenannten kosmologischen Konstanten, ist das Empirieproblem bisher selten artikuliert worden. Will man kosmologische Aussagen innerhalb der Physik machen, kann man sich nicht auf die Metagalaxis beschränken und ansonsten einen flachen Raum postulieren. Mit einem flachen und aktual unendlich ausgedehnten Ortsraum werden alle zugehörigen Weltmodelle physikalisch wertlos. In einem solchen Fall bezieht sich nämlich alle Empirie trivialerweise auf Null Prozent des Ganzen und das ist – für die Physik jedenfalls – unzureichend.

Wenn man also Kosmologie mit einer sinnvollen Anbindung an die empirischen Daten betreiben möchte, ist ein aktual endliches Volumen des kosmischen Raumes eine notwendige Bedingung. Damit können auch all die großartigen Beobachtungen über kosmische Zusammenhänge, die bereits heute einen wesentlichen Teil des Ganzen erfassen, für die physikalische Kosmologie bedeutsam werden.

Diese Bedingung eines endlichen Volumens des Ortsraumes kann auch nicht über ein hypothetisches „kosmologisches Prinzip" kompensiert werden, welches postuliert, dass ein unendlicher Kosmos überall so ist wie der zufällige Teil, den wir davon beobachten.

### Quantentheorie ist universell

Im Zusammenhang mit kosmologischen Fragestellungen wird weiterhin ein weitverbreitetes Missverständnis über Quantentheorie als „Theorie der Mikrophysik" relevant.

Gegenwärtig besteht bezüglich kosmologischer Fragen ein Konsens der Physiker darüber, dass Quantenphänomene dann unvermeidbar werden, wenn man sich der Umgebung des Urknalls und damit winzigen Abständen sowie extremen Dichten und Temperaturen nähert. Dahinter steht die Vorstellung, die Quantentheorie sei „eine Theorie für das Kleine". Selbstverständlich ist es zutreffend, dass „im Kleinen" ohne Quantentheorie nichts verstanden werden kann. Der Grund hierfür liegt aber darin, dass die Quantentheorie als „Physik des Genauen"[4] im Kleinen ersichtlich unverzichtbar ist. Im Großen gibt es hingegen viele Problemstellungen, bei denen die Genauigkeit der Quantentheorie nicht zwingend notwendig ist und die daher ziemlich gut mit der weniger genauen klassischen Physik behandelt werden können. (Diese erlaubt wegen des mit ihr verbundenen Ausschließens der Beziehungsstruktur der Wirklichkeit allerdings gerade deswegen eine exaktere mathematische Gestalt mit einer deterministischen Struktur für die Fakten. Quantentheorie hingegen determiniert lediglich Möglichkeiten.)

Allerdings kann auch im „Großen" die Quantentheorie nicht immer außer Acht gelassen werden – aber auch nicht immer „rezeptförmig" angewendet werden. Ein Beispiel dafür liefert in der Kosmologie der kosmologische Term $\Lambda$ in den Einsteinschen Gleichungen, der zumeist als „Konstante" aufgefasst wird. Einen sehr lesenswerten Überblick über die damit zusammenhängenden Probleme findet man bei Schommers.[5] In der späteren Entwicklung der Physik wurde $\Lambda$ mit der „Energie des Vakuums" in Verbindung gebracht, denn ein solcher Term verkörpert Eigenschaften des Energie-Impuls-Tensors des Minkowski-Raumes. Wenn man aber mit quantenfeldtheoretischen Überlegungen seine Größe berechnet – was im Minkowskiraum im Gegensatz zum Kosmos sinnvoll ist –, so wird das Ergebnis um sehr viele Größenordnungen falsch. Das ist wenig verwunderlich, denn eine solche Konstante ist aus quantenphysikalischer Sicht äußerst fragwürdig. Alle empirischen Daten verweisen auf eine Expansion des kosmischen Raumes. Und zu den ersten Erfahrungen, die man mit der Quantentheorie macht, gehört, dass die Grundzustandsenergie eines Systems – seine „Vakuumenergie" – von dessen räumlicher Ausdehnung abhängt. Sie kann also keineswegs eine davon unabhängige Konstante sein. Wie unten gezeigt wird, lässt sich das Problem mit quanteninformationstheoretischen Überlegungen lösen.

### Ein kosmologisches Modell aus einer abstrakten Quantentheorie

Wenn man also eine Verbindung zwischen Quantentheorie und den Phänomenen der Gravitation sucht, dann erscheint es keineswegs zwingend, von den Einsteinschen Gleichungen als Basis

---

[4] siehe z.B. Görnitz, 1999
[5] Schommers, 2008

auszugehen, die dann zu quantisieren wären.

Als Alternative bietet sich an, aus einer Theorie abstrakter Quanteninformation ein kosmologisches Modell abzuleiten. Für dieses Modell ergeben sich aus plausiblen thermodynamischen Annahmen globale Bedingungen für Energiedichte und Druck. Ein daraus gebildeter Energie-Impuls-Tensor zeigt zwar keine offenbaren Beziehungen zur Metrik des kosmologischen Modells. Geht man aber von der Metrik zu deren Ableitungen über, so wird der sich dabei ergebende Einstein-Tensor proportional zum gefundenen Energie-Impuls-Tensor.

**Die nun nahe liegende Forderung, dass diese Beziehung zwischen den beiden Tensoren auch für den Fall von Abweichungen von der homogenen Verteilung von Energiedichte und Druck erhalten bleiben soll, hat die Gültigkeit der Allgemeinen Relativitätstheorie als lokale Näherung zur Folge.**

In einer Reihe von Arbeiten wurde ein solches mit quantentheoretischen Überlegungen begründetes kosmologisches Modell vorgestellt.[6] Es beschreibt einen geschlossenen Kosmos, der sich mit Lichtgeschwindigkeit ausdehnt. Allerdings war diese Modell seiner Zeit zu weit voraus; denn damals vertrat die kosmologische Folklore noch die Meinung, dass ein geschlossener Kosmos rekollabiert und sich keineswegs dauerhaft nur ausdehnen kann. Anschließend wurde dann die Vorstellung populär, dass der Kosmos nicht geschlossen, sondern flach sei. Daher wurde dem hier vorzustellenden Modell bisher keine größere Aufmerksamkeit zuteil, obwohl es nach einigen Jahren sogar im Lehrbuch auftauchte.[7]

In der Zwischenzeit ist Quanteninformation keineswegs mehr lediglich eine Überlegung von theoretischen Außenseitern, sondern in einer wachsenden Zahl von höchst erfolgreichen Experimenten werden Eigenschaften von Quanteninformation deutlich gemacht.[8]

Heute zeigt sich, dass das hier vorgestellte kosmologische Modell nicht nur wesentliche Probleme älterer Modelle beseitigt, sondern dass es jetzt auch besser zu den empirischen Ergebnissen passt[9] als zu den Daten, die früher als zutreffend angesehen wurden.

Gestützt werden die vorliegenden Überlegungen durch eine wichtige Arbeit von Jacobson.[10] Dieser zeigt, wie die Einsteinschen Gleichungen aus der Thermodynamik der Schwarzen Löcher hergeleitet werden können. Er schreibt:

---

[6] Görnitz, 1988 a,b, Görnitz, Ruhnau, 1989
[7] Goenner, 1994, S. 87
[8] siehe z. B. Arbeiten Zeilinger, Weinfurtner und vielen anderen.
[9] Görnitz, 2006, Görnitz & Görnitz, 2008, S. 148
[10] Jacobson, 1995

"The four laws of black hole mechanics, which are analogous to those of thermodynamics, were originally derived from the classical Einstein equation. With the discovery of the quantum Hawking radiation, it became clear that the analogy is, in fact, an identity. How did classical general relativity know that the horizon area would turn out to be a form of entropy, and that surface gravity is a temperature? In this Letter I will answer that question by turning the logic around and deriving the Einstein equation from the proportionality of entropy and the horizon area together with the fundamental relation $\delta Q = T\, dS$ connecting heat $Q$, entropy $S$, and temperature $T$. Viewed in this way, the Einstein equation is an equation of state. It is born in the thermodynamic limit as a relation between thermodynamic variables, and its validity is seen to depend on the existence of local equilibrium conditions. This perspective suggests that it may be no more appropriate to quantize the Einstein equation than it would be to quantize the wave equation for sound in air."

Jacobsons Herleitung beruht auf der Thermodynamik und der Hawking-Strahlung. Nachfolgend soll deutlich gemacht werden, dass die Hawking-Strahlung ein Ausdruck für die Entropie der Raumzeit, d.h. für die der Raumzeit insgesamt zugrunde liegenden Quanteninformation ist. Damit wird es möglich, nicht mit der Hawking-Strahlung zu starten, sondern stattdessen mit einer noch fundamentaleren abstrakten Quanteninformation.

Seit den 50er Jahren vertrat C.F. v. Weizsäcker die Ansicht, dass die Physik mit ihren Objekten auf eine abstrakte Quanteninformation – Ur-Alternativen – gegründet werden kann.[11] Dass dies eine vernünftige Annahme ist, wurde spätestens deutlich, als von Görnitz, Graudenz und v. Weizsäcker gezeigt wurde, dass Quantenteilchen aus Qubits erzeugt werden können. Als Objekte der relativistischen Quantenmechanik werden sie durch irreduzible Darstellungen der Poincarégruppe erfasst. Diese können durch Para-Bose-Erzeuger und -Vernichter von Quantenbits im Rahmen einer „zweiten Quantisierung" beschrieben werden.[12]

Weizsäcker postulierte, dass die Dreidimensionalität des Raumes auf der SU(2)-Symmetrie der Qubits beruhen würde. Dies ist von Drieschner[13] weiter ausgearbeitet worden. Allerdings standen diese damals verwendeten Modelle im Widerspruch zur Allgemeinen Relativitätstheorie, was ihre Akzeptanz stark behindert hatte.

Da Weizsäcker die These aufgestellt hatte, „Ein »absoluter« Begriff der Information hat

---

[11] Weizsäcker, 1955, 1958, 1971, Scheibe et al., 1958
[12] Görnitz, Graudenz, v. Weizsäcker, 1992, Görnitz, Schomäcker, 1996
[13] Drieschner, 1979

keinen Sinn"[14], konnte seinen Ur-Alternativen keine absolute Größe zugesprochen werden. Es wurde daher nötig, über Weizsäckers Diktum hinauszugehen und den Begriff der Information so abstrakt zu fassen, dass dabei weder Sender noch Empfänger und vor allem auch keine Bedeutung mitgedacht werden. Nur mit einer solchen weiteren Abstraktion kann Quanteninformation zu einer absoluten Größe werden und damit äquivalent zu Materie und Energie.

Unter Zuhilfenahme gruppentheoretischer und thermodynamischer Argumente über Schwarze Löcher konnte von Görnitz[15] ein kosmologisches Modell hergeleitet werden, welches nicht im Widerspruch zur Allgemeinen Relativitätstheorie steht. Erst mit dieser Anbindung an Kosmologie und Schwarze Löcher wurde eine absolute Quanteninformation definierbar, die wegen ihrer Bedeutungsfreiheit mit einem neuen Namen bedacht wurde – Protyposis – um die zu einengende Assoziation Information = Bedeutung aufzubrechen.

### Postulate für eine Kosmologie

Nachfolgend wird gezeigt, dass in einer Umkehrung der Argumentationsreihenfolge mit vier physikalisch sinnvollen Annahmen aus einer abstrakten und vorerst bedeutungsfreien Quanteninformation – Protyposis – ein kosmologisches Modell abgeleitet werden kann, aus dem seinerseits die Allgemeine Relativitätstheorie folgt.

1. *Abstrakte Quanteninformation – Protyposis – bildet die Basis der Physik.*
2. *Die Energie eines Quantensystem ist proportional zum Inversen seiner Wellenlänge*
3. *Für ein abgeschlossenes System gilt der erste Hauptsatz der Thermodynamik: $dU + p\,dV = 0$*
4. *Es existiert eine ausgezeichnete Geschwindigkeit: c*

### Einführung des Ortsraumes

Der Zustandsraum eine Qubits ist der zweidimensionale komplexe Raum $\mathbb{C}^2$. Die Symmetriegruppe, die den Betrag des Skalarprodukts in diesem Hilbertraumes invariant lässt, umfasst im wesentliche die Gruppe SU(2), dazu die Phasentransformationen der U(1) und die Komplexkonjugation.

---

[14] v. Weizsäcker, 1985, S. 172
[15] Görnitz, 1988

Die SU(2) als maximaler homogener Raum der Gruppe ist geometrisch eine $\mathcal{S}^3$, die dreidimensionale Oberfläche einer vierdimensionalen Kugel. Der Hilbertraum der quadratintegrablen Funktionen auf dieser $\mathcal{S}^3$ trägt die reguläre Darstellung der Gruppe, in der jede irreduzible Darstellung enthalten ist. Eine zweidimensionale Darstellung, die alle Zustände eine Qubits repräsentiert, wird von solchen Funktionen auf der $\mathcal{S}^3$ aufgespannt, die lediglich eine einzige Knotenfläche enthalten. Werden viele Qubits quantentheoretisch kombiniert, d.h. das Tensorprodukt ihrer Zustandsräume gebildet, dann werden in diesem Produkt sehr viel besser lokalisierte Zustände erreichbar.

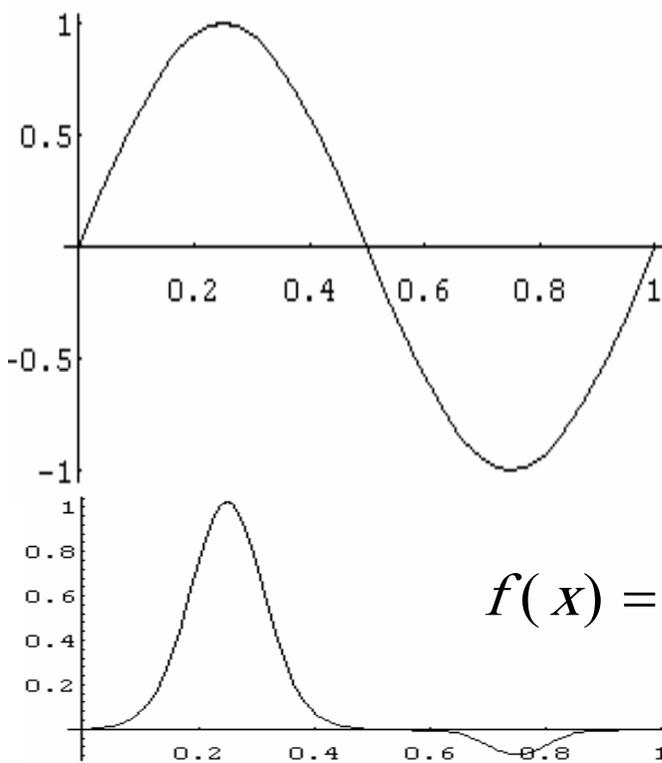

$$f(x) = \sin(2\pi x)$$

$$f(x) = \frac{1}{44} \sum_{n=1}^{9} n(\sin 2\pi x)^n$$

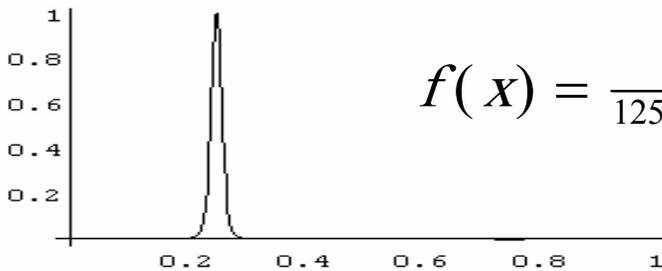

$$f(x) = \frac{1}{125000} \sum_{n=1}^{501} n(\sin 2\pi x)^n$$

**Abb 1: Hohe Potenzen des Sinus erlauben starke Lokalisation**

Ein eindimensionales Beispiel, wie aus vielem Ausgedehnten etwas Lokalisiertes gebildet werden kann, soll der Einfachheit halber am Sinus verdeutlicht werden, der das Einheitsintervall ebenfalls in lediglich zwei Teile zerlegt. Mit dem Produkt von vielen Sinus-Funktionen werden auch stark lokalisiert Zustände erreichbar. (Abb. 1)

Die durch die Quantentheorie ermöglichte Erstellung von lokalisierten Erscheinungen aus extrem ausgedehnten Ausgangsteilen ist ein zentraler Gesichtspunkt der hier vorgestellten Theorie. Hier wird nicht mehr postuliert, dass das räumlich Kleine zugleich auch das Einfache sein muss. Es zeigt sich im Gegenteil, dass das Einfachste, das sich denken lässt, ein Qubit, über den Kosmos maximal ausgedehnt ist.

Zugleich ist der Unterschied zu einem Quantenfeld in extremer Weise ausgeprägt. Für ein Quantenfeld kann postuliert werden, dass im Prinzip für jeden Raumpunkt ein Wert der Feldstärke beliebig vorgegeben werden kann. Für die Funktion, welche den Zustand des Qubits repräsentiert, gibt es derartige Freiheiten nicht. Die Lage eines Maximums legt die gesamte Funktion im ganzen Raum fest. Dieser Unterschied lässt auch leichter verstehbar werden, dass der feldtheoretisch berechnete extrem fehlerhafte Wert für den kosmologischen Term in vorliegenden Zugang nicht auftritt.

Wenn im Kosmos N Qubits vorhanden sind, so ist die erste Frage, mit welchen Wellenlängen auf der $\mathscr{S}^3$ dann aus physikalischen Gründen zu rechnen ist. Eine Antwort ergibt sich aus der Ausreduktion des Tensorproduktes von N zweidimensionalen Darstellungen $D_{1/2}$.[16] Für die Clebsch-Gordan-Reihe ergibt sich mit der Festlegung:

$$|N/2| = k \text{ für } N = 2k \text{ oder } N = 2k+1.$$

$$D_{\frac{1}{2}}^{\otimes N} = \bigoplus_{j=0}^{|N/2|} \frac{N!(N+1-2j)}{(N+1-j)!\,j!} D_{|N/2|-j}$$

Der Faktor der jeweiligen Multiplizitäten

$$f(j) = \frac{N!(N+1-2j)}{(N+1-j)!\,j!}$$

---

[16] Görnitz, 1988

bietet ein Maß dafür, wie viele Funktionen mit welchen Wellenlängen vorkommen können. Die Multiplizität der größten Wellenlängen, die zu $D_0$ bzw. $D_{1/2}$ und damit zu $j=|N/2|$ gehören, sind von der Größenordnung

$$f(N/2) = \mathcal{O}\,(2^N N^{3/2})$$

Zu kürzeren Wellenlängen hin wächst die Multiplizität an und erreicht bei

$$j \approx \tfrac{1}{2}(N - \sqrt{N})$$

ihr Maximum in der Größenordnung

$$f\left(\tfrac{1}{2}(N - \sqrt{N})\right) = O\!\left(2^N N^{-1}\right) = \sqrt{N} \cdot O\!\left(2^N N^{-3/2}\right)$$

Nach diesem im wesentlichen linearen Anwachsen folgt ein exponentieller Abfall der Multiplizität. Das bedeutet, dass man Zustände, die zu wesentlich kleineren Wellenlängen gehören, in der physikalischen Praxis nicht finden wird.

Bezeichne R den Krümmungsradius der $\mathcal{S}^3$, so folgt eine kleinste, physikalisch noch realisierbare Länge von der Größenordnung

$$\lambda_0 = \frac{R}{\sqrt{N}}$$

Die Zustandsfunktion, die ein Qubit repräsentiert, hat nach diesen Überlegungen eine Wellenlänge der Größenordnung $\sqrt{N}\cdot\lambda_0$. Nach der hier vorausgesetzten Prämisse folgt daraus, dass einem solchen trägerfreien Qubit eine Energie der Größenordnung $\cdot 1/\sqrt{N}$ zuzuordnen ist. Die Gesamtenergie *U* der *N* Qubits ist dann von der Größenordnung

$$N/\sqrt{N} = \sqrt{N}$$

Da das Volumen der $\mathcal{S}^3$ von der Größenordnung *R³* ist, so folgt für die Energiedichte *μ* ein Wert proportional zu $\sqrt{N}/R^3$, d.h. proportional zu *1/R²*,

Nach einer weiteren vorausgesetzten Prämisse gilt der erste Hauptsatz

$$dU + p\,dV = 0$$

und damit

$$dR + p\,3\,R^2\,dR = 0$$

oder

$$p = -1/3\, R^2$$

Damit ergibt sich ein negativer Druck. Daraus folgt, dass ein solches System nicht statisch sein kann.[17] Allerdings erfüllt dieser Druck alle die Bedingungen, die man aus physikalischen Gründen an die Beziehungen zwischen Energiedichte und Druck stellen muss.[18]

Für die Festlegung der Expansion der kosmischen $\mathcal{S}^3$ wird die Prämisse einer ausgezeichneten Geschwindigkeit benutzt. Die zu verwendende Zeiteinheit ergibt sich damit so, dass

$$R = c\, T$$

gilt.

So ist aus quantentheoretischen Überlegungen in Verbindung mit den obigen vier Prämissen ein kosmologisches Model hergeleitet worden, ohne dass dabei bereits auf eine Theorie der Gravitation Bezug genommen worden wäre.

Der sich ergebende Kosmos ist eine homogene und isotrope $\mathcal{S}^3$, die mit Lichtgeschwindigkeit expandiert.

Der gefundene Energie-Impuls-Tensor hat die Gestalt

$$T^i_{\,k} = diag\,(\mu,\, p,\, p,\, p) = (1/R^2,\, -1/3R^2,\, -1/3R^2,\, -1/3R^2)$$

und die Metrik ist

$$ds^2 = dt^2 - (ct)^2\,[(1-r^2)^{-1}\, dr^2 + r^2\, d\Omega^2)]$$

Die so hergeleiteten Bedingungen für kosmischen Druck und Energiedichte erfüllen alle aus physikalischen Gründen zu fordernden Bedingungen, wie sie z.B. bei Hawking und Ellis hergeleitet werden.[19]

Diese sind erstens die schwache Energiebedingung

$$\mu \geq 0 \quad und \quad \mu + p \geq 0$$

die zur Folge hat, dass jeder Beobachter **in seinem Ruhsystem immer eine positive Energiedichte** messen wird.

Weiter gibt es die dominante Energiebedingung

---

[17] Landau, Lifschitz, 1971, S. 47
[18] Hawking, Ellis, 1973
[19] Hawking, Ellis, 1973

$$\mu \geq 0 \quad und \quad \mu \geq p \geq -\mu$$

aus welcher folgt, dass Geschwindigkeit des Energieflusses nie größer als die Lichtgeschwindigkeit werden kann bzw. dass die **Schallgeschwindigkeit stets kleiner als Lichtgeschwindigkeit** bleibt.

Schließlich hat die starke Energiebedingung

$$\mu + 3p \geq 0 \quad \mu + p \geq 0$$

die Konsequenz, dass die **Gravitation stets anziehend** wirkt.

### Die Lösung der kosmologischen Probleme

Neben dem bereits von mir erwähnten Empirie-Problem werden bisher in der Kosmologie noch eine Reihe weiterer Probleme aufgeführt. Zu ihnen gehören das Horizontproblem, das Problem der „kosmologischen Konstanten", das Flachheitsproblem und die „Dunkle Energie".

Unter dem Horizontproblem wird die Erklärungsnotwendigkeit dafür gesucht, dass die Hintergrundstrahlung aus allen Richtungen im Promillebereich identisch ist, obwohl nach vielen kosmologischen Modellen zuvor zwischen diesen Bereichen nie ein kausaler Kontakt möglich gewesen wäre.

Die bisher zur Lösung erfundene Inflation verletzt mit ihrer Zustandsgleichung $\rho = -p$ Bedingungen, die man aus vernünftigen Gründen an eine physikalische Substanz stellen muss.[20] Das kosmologische Modell, das aus der Protyposis folgt, besitzt diese Probleme nicht. Da sich nach ihm der Kosmos stets mit Lichtgeschwindigkeit ausdehnt, existieren in ihm keine Bereiche, die zuvor in keinem kausalen Kontakt gewesen waren.

Das Problem der „kosmologischen Konstanten" besteht darin, einen Grund für die Existenz einer Größe zu finden, die in Planckschen Einheiten den Wert $10^{-122}$ besitzt, die also mit 122 Nullen beginnt, aber nicht Null ist. Einstein hatte sie in seine Gleichung eingefügt, um ein statisches Universum zu erzwingen. In den 1980er Jahren hatte man sich dafür entschieden, sie gleich Null zu setzen. In den letzten Jahren jedoch zeigte es sich, dass ein solcher Verzicht zu Widersprüchen mit den Beobachtungsdaten führt.

---

[20] Hawking, Ellis, 1973

In den kosmologischen Modellen hatte man als wesentliche Komponenten eine druckfreie Materie (Staub) und die masselose Strahlung (Licht) neben Λ eingeführt.

Ein Qubit der Protyposis ist eine extrem nichtlokale Struktur. Als Objekte bezeichnet man in der Physik solche Strukturen, die lokalisiert werden können und deshalb auch in Raum und Zeit bewegt werden können. Dabei sollen sie lediglich ihren Zustand verändern, nicht aber sich selbst. In einer exakten mathematischen Form lässt sich dies im Minkowski-Raum fassen. In ihm ist sauber definiert, was eine Teilchen ist, nämlich eine Struktur, deren Zustände durch eine irreduzible Darstellung der Poincarégruppe beschrieben werden. Dann ergeben sich bekanntlich masselose und massive Objekte, die sich noch in ihrem Spin unterscheiden können und die in der Physik gemeinhin alle als „Teilchen" bezeichnet werden.

Die Protyposis kann sich teilweise zu Materie und Strahlung ausformen. Wenn man ihren Energie-Impuls-Tensor in die Tensoren von Materie und Strahlung zerlegt, so ergibt sich die Notwendigkeit, einen weiteren Tensor einzuführen, der die Struktur von Λ besitzt – ohne allerdings zeitlich konstant sein zu müssen. Der mit δ bezeichnete nichtlokalisierte Anteil der Protyposis wird sich global als „Dunkle Energie" in der Kosmologie bemerkbar machen, eine zusätzliche Ausformung in „Teilchen" ist dabei nicht notwendig.

$$T^i_k = {}_{(Staub)}T^i_k + {}_{(Licht)}T^i_k + {}_{(Vakuum)}T^i_k + {}_{(dunkle\,Energie)}T^i_k$$

$$\begin{bmatrix} \mu & & & \\ & \frac{\mu}{3} & & \\ & & \frac{\mu}{3} & \\ & & & \frac{\mu}{3} \end{bmatrix} = \begin{bmatrix} \mu_{matter} & & & \\ & 0 & & \\ & & 0 & \\ & & & 0 \end{bmatrix} + \begin{bmatrix} \mu_{light} & & & \\ & \frac{-\mu_{light}}{3} & & \\ & & \frac{-\mu_{light}}{3} & \\ & & & \frac{-\mu_{light}}{3} \end{bmatrix} + \begin{bmatrix} \lambda & & & \\ & \lambda & & \\ & & \lambda & \\ & & & \lambda \end{bmatrix} + \begin{bmatrix} \delta\mu & & & \\ & \delta\frac{\mu}{3} & & \\ & & \delta\frac{\mu}{3} & \\ & & & \delta\frac{\mu}{3} \end{bmatrix}$$

Mit ω werde das Verhältnis der Energiedichte vom Materie und Licht zu der Energiedichte des Vakuums bezeichnet:

$$\omega = (\mu_{(matter)} + \mu_{(light)}) / \lambda$$

Dann erhalten wir[21]

$$\lambda = (1-\delta)\,\mu / (\omega + 1)$$

$$\mu_{(light)} = (1-\delta)\,\mu\,(2 - \omega)/(\omega + 1)$$

$$\mu_{(matter)} = (1-\delta)\,2\mu(\omega - 1)/(\omega + 1)$$

---

[21] Görnitz, 1988

wegen   $\mu > 0$,   $\mu_{(matter)} > 0$,   $\mu_{(light)} > 0$   folgt   $2 \geq \omega \geq 1$

Mit dieser Betrachtung wird das Problem des „Kosmologischen Terms" gelöst. Es zeigt sich, dass die Größe $\lambda$, die die Eigenschaften einer Energiedichte des Vakuums besitzt, von der gleichen Größenordnung sein muss wie die Energiedichte vom Materie und Licht und keineswegs – wie es sich aus quantenfeldtheoretischen Rechnungen ergeben hatte – um sehr viele Größenordnungen größer.

Das „Flachheitsproblem" löst sich dahingehend auf, dass eine $\mathcal{S}^3$, die seit 14 Mrd. Jahren mit Lichtgeschwindigkeit expandiert, lokal durchaus sehr flach erscheinen muss.

### Vergleich mit den Beobachtungen

Wenn diese Metrik mit den Beobachtungen verglichen wird, so sind die modernsten Messergebnisse sehr ermutigend. Das Protyposis-Modell fordert für den Hubble-Parameter $H$, dass *$HT = 1$.* Die Ergebnisse der Supernova Ia-Daten zeigen, dass *$H_0\, T_0 = 0.96 \pm 0.04$.*

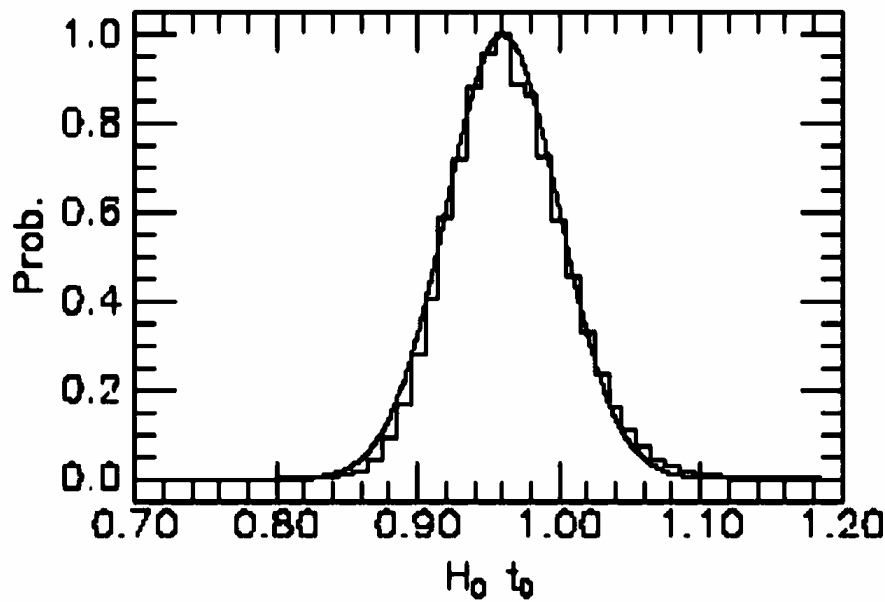

**Abb. 2: Beziehung zwischen Hubble-Parameter und Weltalter**[22]

Gegenwärtig wird das Bild einer Kosmologie bevorzugt, nach welcher der kosmische Raum beschleunigt expandiert. Wenn die modernen Supernova-Daten in Betracht gezogen werden, so zeigt sich ebenfalls ein positiver Bezug zum hier vorgestellten Modell.

---

[22] Tonry et al, (2003), Fig. 15

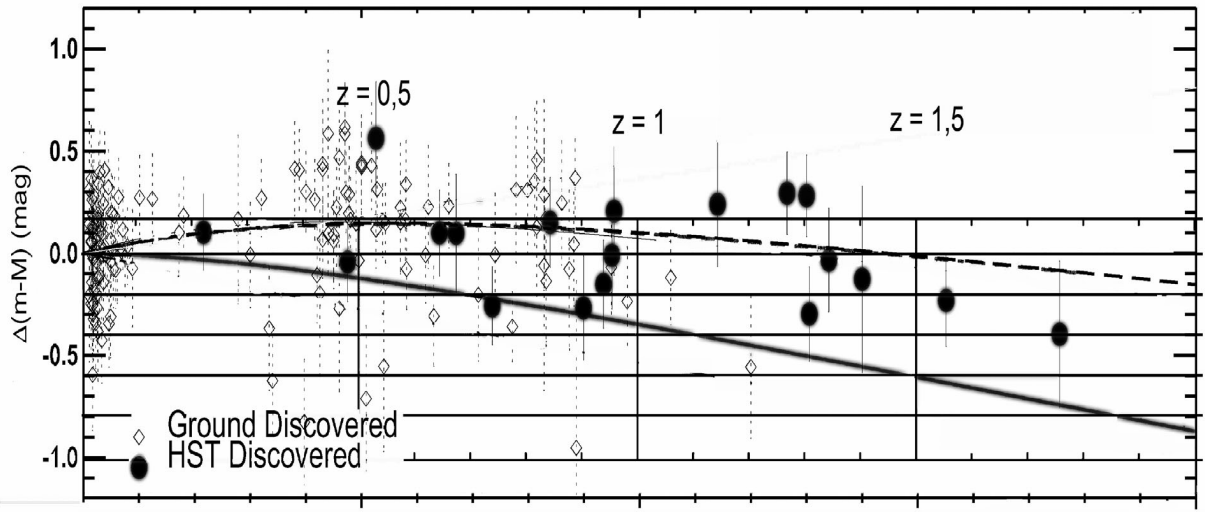

**Abb. 3:** Vergleich der Supernova-Daten[23] mit dem Modell eines beschleunigt expandierenden Kosmos (- - -) und mit einem mit Lichtgeschwindigkeit expandierenden Kosmos (——)[24]

Mit den vorgelegten Überlegungen ist zu erwarten, dass sich die Tendenz der Daten verstärken wird, sich für größere Entfernungen, d.h. bei noch höheren Werten von z, der unteren Kurve eines konstant expandierenden Kosmos anzunähern. Diese Erwartung stützt sich auch auf das Argument von Aurich, Lustig und Steiner[25]: „but there remains a strange discrepancy at large scales as first observed by COBE[26] and later substantiated by WMAP[27]." und weiter: „The suppression of the CMB anisotropy at large scales respectively low multipoles can be explained if the Universe is finite."

Für eine Klärung des Zusammenhanges der Protyposis-Kosmologie mit der Struktur der Hintergrundstrahlung sind weitere Untersuchungen in Vorbereitung, denn Aurich, Lustig und Steiner schreiben auch, dass mit den von ihnen verwendeten Parametern die durchgeführten Rechnungen für die einfach zusammenhängende $\mathcal{S}^3$ noch nicht die nötigen Ergebnisse zur Folge hatten. Diese Autoren haben sich deshalb komplizierteren Topologien zugewendet, d.h. mehrfach zusammenhängende Räume untersucht.

---

[23] Riess et. al. , 2004, Fig. 7
[24] Görnitz, 2006, S. 163 ff
[25] Aurich, Lustig, Steiner, 2005
[26] Hinshaw et al., 1996
[27] Bennett et al., 2003

### Begründung der Allgemeinen Relativitätstheorie aus der Protyposis

Die Protyposis hat zu einem kosmologischen Modell geführt, in dem eine Hintergrundmetrik ausgezeichnet erscheint. Bereits Dirac hat darauf verwiesen, dass von einer solchen ausgezeichneten Hintergrundmetrik nichts zu bemerken ist, solange man in einem „Einsteinschen Fahrstuhl" ohne Fenster verbleibt. Dies ändert sich, wenn man ein Fenster zum Kosmos öffnet und auch die Mikrowellen-Hintergrundstrahlung betrachtet.[28] Die Gleichberechtigung aller Bezugssysteme ist daher als eine lokale Eigenschaft der Beschreibung von Systemen anzusehen, die beliebigen Beschleunigungen unterliegen. Es ist jedoch nicht notwendig, diese Eigenschaft auch für die Kosmologie zu fordern.

Wenn man die Metrik aus dem Protyposis-Modell mit seinem Energie-Impuls-Tensor vergleicht, so ist auf den ersten Blick keine Beziehung zu erkennen. Wenn man aber für diese Friedman-Robertson-Walker-Metrik den Einstein-Tensor $G^i_k$ bildet, so erkennt man sofort, dass $G^i_k$ dem gefundenen Energie-Impuls-Tensor proportional ist.

**Fordert man nun, dass diese Proportionalität auch für die lokalen Störungen der homogenen und isotropen Hintergrundmetrik weiterhin erhalten bleibt, so wird damit die Allgemeine Relativitätstheorie als Beschreibung von lokalen Inhomogenitäten dieser Kosmologie erhalten.**

### Literatur

---

[28] Dirac, 1980